\documentclass[aps,prb,twocolumn,superscriptaddress,showpacs,amsmath,floatfix,citeautoscript]{revtex4}

\usepackage{amssymb}
\usepackage{amsmath}
\usepackage{colordvi}
\usepackage[colorlinks]{hyperref}
\usepackage{amsthm}
\usepackage{subeqnarray}
\usepackage{cases}
\usepackage{enumerate}
\usepackage{graphicx}
\usepackage{epsfig}
\usepackage{epstopdf}
\usepackage{subfigure}
\usepackage{color}
\usepackage{ulem}

\def\mc{\multicolumn}

\def\avg(#1){\langle#1\rangle}

\def\be{\begin{equation}}
\def\ee{\end{equation}}
\def\bea{\begin{eqnarray}}
\def\eea{\end{eqnarray}}

\begin{document}
\title{Gradient optimization of fermionic projected entangled pair states on directed lattices}

\author{Shao-Jun Dong}.
\affiliation{Key Laboratory of Quantum Information, University of Science and
  Technology of China, Hefei, 230026, China}
\affiliation{Synergetic Innovation Center of Quantum Information and Quantum
  Physics, University of Science and Technology of China, Hefei, 230026, China}

\author{Chao Wang}
\affiliation{Key Laboratory of Quantum Information, University of Science and
  Technology of China, Hefei, China}
\affiliation{Synergetic Innovation Center of Quantum Information and Quantum
  Physics, University of Science and Technology of China, Hefei, 230026, China}

\author{Yongjian Han}
\email{smhan@ustc.edu.cn}
\affiliation{Key Laboratory of Quantum Information, University of Science and
  Technology of China, Hefei, China}
\affiliation{Synergetic Innovation Center of Quantum Information and Quantum
  Physics, University of Science and Technology of China, Hefei, 230026, China}

\author{Guang-can Guo}
\affiliation{Key Laboratory of Quantum Information, University of Science and
  Technology of China, Hefei, China}
\affiliation{Synergetic Innovation Center of Quantum Information and Quantum
  Physics, University of Science and Technology of China, Hefei, 230026, China}

\author{Lixin He}
\email{helx@ustc.edu.cn}
\affiliation{Key Laboratory of Quantum Information, University of Science and
  Technology of China, Hefei, China}
\affiliation{Synergetic Innovation Center of Quantum Information and Quantum
  Physics, University of Science and Technology of China, Hefei, 230026, China}

\begin{abstract}
The recently developed stochastic gradient method combined with Monte Carlo sampling techniques
[PRB {\bf 95}, 195154 (2017)] offers a low scaling and accurate method to optimize
the projected entangled pair states (PEPS).
We extended this method to the fermionic PEPS (fPEPS).
To simplify the implementation, we introduce a Fermi arrow notation to specify
the order of the fermion operators in the virtual entangled EPR pairs. By defining some
local operation rules associated with the Fermi arrows,
one can implement fPEPS algorithms very similar to that of standard PEPS.
We benchmark the method for the interacting spin-less fermion models, and the t-J models.
The numerical calculations show that the gradient optimization greatly improves the results of
simple update method. Furthermore, very large virtual bond dimensions ($D$)
and truncation dimensions ($D_c$) are necessary to converge the results of these models.
The method therefore offer a powerful tool to simulate fermion systems because it has much lower scaling than
the direct contraction methods.
\end{abstract}
%\pacs{xxxxxx}
\maketitle

\section{introduction}

Interacting quantum many-body systems pose some of the most exciting open problems in physics. Particularly,
fermion systems are central to many of the most fascinating effects in condensed matter physics,
such as high-temperature superconductivity,\cite{Lee2006} the fractional quantum Hall effect,\cite{Stormer1999}  and Mott insulator transitions \cite{EDWARDS1968,Imada1998}. The simulation of the strongly correlated fermion system plays the critical role to understand these system and is also one of the most challenging problems in condensed matter physics.

The Quantum Monte Carlo (QMC)\cite{Foulkes2001} method as one of the leading methods in studying many-body physics has achieved great success in bosonic and spin systems since its first proposed. However, except in some special cases,~\cite{Li2018}
%e.g., one-dimensional \cite{} and half-filling systems \cite{},
the fermion systems are extremely difficult to treat using QMC simulations\cite{Loh1990,Troyer05} because of the notorious ``sign problems''.

Recently, the methods based on tensor network states (TNS), especially the projected entangled states (PEPS) \cite{Schollwoeck2011,Garcia2006,Verstraete2008,Xiang2008,Vidal2008,Verstraete2004,Cirac2010,Verstraete06} have shown their power on simulation of the strongly correlated many-particle systems.
The PEPS is sign-problem free and has achieved great successes in studying the frustrated spin models \cite{wang16,Vidal2007,Alexander2012}. The PEPS method has been extended to study fermion models (namely fPEPS) by different approaches \cite{Barthel2009,Corboz2010,Corboz2011,Gu2010,Gu2013,Kraus2010}. Apperently, the fPEPS are more complicated than PEPS because of the anti-commutation properties of the fermion operators. In addition, fermion systems are highly frustrated. It has been proven that the entanglements of
the ground states of some fermion systems are beyond the area law \cite{Hastings2007,Wolf2006}. Therefore, to faithfully simulate such models, it usually requires very large bond dimensions ($D$).
Furthermore, it has been shown that the imaginary time evolution with simple update \cite{Xiang2008} method may have large errors because the environment effects are oversimplified.
To exactly consider the environment, the traditional methods, e.g., the full update method \cite{Lubasch2014,Michael2014},
suffer from extremely high computational scaling to the bond dimensions.
This problem is more serious for the fermion models when large $D$ is required.
We note that the recently developed infinite PEPS (iPEPS)
with full update method has achieved great success,\cite{Corboz2014,Corboz2010}
by making use of
the translation symmetry, which may greatly reduce the number of independent tensors. However, not all systems have such symmetry, e.g., defects, disorders and systems with spontaneous symmetry broken, etc.
In these cases, the finite PEPS method is essential.

The recently developed Monte Carlo sampling techniques for PEPS can greatly reduce the computational scaling\cite{Sandvik2007,Schuch2008,Cirac2010,Banuls2017,Dong2017,Liu2017}. By combining with stochastic gradient
optimization (GO) method, one can achieve great precision in obtaining the ground states.~\cite{Liu2017,He2018}
In this work, we extended the stochastic gradient method~\cite{Liu2017,He2018}
to optimize the fPEPS wave functions for fermion systems.
To simplify the implementation of the fPEPS algorithms,
we introduce a ``Fermi arrow'' notation to specify
the order of the fermion operators in the entangled EPR pairs.
With this notation and some {\it local} operation rules associated with the Fermi arrows,
we can greatly simplify the implementation of the stochastic gradient optimization
method (and other methods) for fPEPS. We implement this local operation rules for fPEPS in our recently developed TNSpack \cite{Dong2017libTNSP},
in which the anti-commutation properties of the fermion operators are automatically taken account of. Therefore, one can implement fPEPS algorithm very  similar to that of the standard PEPS without worry too much about the details of the anti-commutation between the fermion operators.

We benchmark the stochastic gradient method for fPEPS on the interacting spin-less fermion models,
and the t-J models. The numerical calculations show that the gradient optimization greatly improves the results of
simple update method. Furthermore, for these models, very large
virtual bond dimensions $D$ and truncation dimensions $D_c$ are necessary to converge the results which is the dominate difficult to simulate the fermion systems.
Therefore the present method is advantageous because it has much lower scaling
than the traditional direct contraction method.

\section{Definition of fPEPS based on directed network}
\label{sec:fpeps}

The definition of the fPEPS\cite{Kraus2010} on a lattice is similar to that of the standard PEPS\cite{Verstraete2004,Verstraete06}.
% We only consider the fPEPS with open boundary condition in this paper.
 Without lose of generality, we take a fermion system on a $L_1 \times L_2$ square lattice as an example,
where the physical dimension of each site is $d$. For each horizontal bond connecting sites $(i,j)$ and $(i,j+1)$, there is a EPR pair, i.e., a Bell type entangled state,
\begin{equation}
 \hat{I}_{h}(i,j)|0\rangle=\sum^{D-1}_{k=0}|k\rangle_{(i,j)_{r}}|k\rangle_{(i,j+1)_{l}} \, ,
\end{equation}
where $|k\rangle_{(i,j)_r}$ and $|k\rangle_{(i,j+1)_l}$ are the fermion states on site $(i,j)$ and site $(i,j+1)$. States $|k\rangle$ are generated as, $|k\rangle=|k_1k_2\cdots k_p\rangle={a_1^{\dag k_1}}{}{a_2^{\dag k_2}}\cdots {a_p^{\dag k_{p}}}|0\rangle$, where $(k_1k_2\cdots k_p$) is the binary representation of $k$ and $|0\rangle$ is the vacuum state.  $a_i$s and $a^\dagger_j$s are the fermion operators that satisfy $\{a_i,a^\dagger_j\}=\delta_{ij}$. For convenience, we denote the state $|k_{(i,j)_r}\rangle={a_{(i,j)_r}^{\dag k}}|0\rangle$.
%EPR state can be rewritten in fermion operator as:
%%
%\begin{equation}
%|I\rangle_{h}(i,j)=\frac{1}{\sqrt{D}}\sum^{D-1}_{k=0}{a_{(i,j)_r}^{\dag}}^k{a_{(i,j+1)_l}^{\dag}}^k|0\rangle\, .
%\end{equation}
%
Similarly, for each vertical bond connecting site $(i,j)$ and $(i+1,j)$, there is also a Bell type entangled state, (in short) $\hat{I}_{v}(i,j)|0\rangle =\sum^{D-1}_{k=0}{a_{(i,j)_d}^{\dag k}}{a_{(i+1,j)_u}^{\dag k}}|0\rangle$. Therefore, a standard virtual mother state of a fPEPS can be defined as,
\begin{equation}
 |\Phi_0\rangle=\Pi^{L_1-1}_{i=1}\Pi^{L_2-1}_{j=1} \hat{I}_{h}(i,j)\hat{I}_{v}(i,j) |0\rangle.
\end{equation}
To define a quantum state in the real physical space, we project $|\Phi_0\rangle$
to the physical space. The projector on site $(i,j)$ is defined as:

\begin{eqnarray}\label{eq:projector}
\hat{P}[i,j] &=&\sum^{d-1}_{\beta=0}\sum^{D-1}_{\beta_1,\beta_2,\beta_3,\beta_4}T_{\beta,\beta_1,\beta_2,\beta_3,\beta_4}[i,j]a^{\dag,\beta}_{(i,j)}\\ \nonumber
&& a_{(i,j),l}^{\beta_1}a_{(i,j),d}^{\beta_2}a_{(i,j),r}^{\beta_3}a_{(i,j),u}^{\beta_4}\, .
\end{eqnarray}
Here, $a^{\dag}_{(i,j)}$ is the creation operator of the physical particle on site $(i,j)$
whereas $a_{(i,j),m}^n$ ($m=l,d,r,u$ and $n=0,1,\cdots,D-1$) are the annihilation operators of the state $|n\rangle_{(i,j),m}$.
%, i.e. $a_{(i,j),m}^n|n\rangle_{(i,j),m}=|0\rangle$ .
The fPEPS is then defined as,
\begin{equation}\label{equ:fPEPS}
|\Phi_{\rm fPEPS}\rangle=\Pi_{i,j}\hat{P}[i,j]|\Phi_0\rangle\, .
\end{equation}

To make the fPEPS well defined, the state $|\Phi_{\rm fPEPS}\rangle$ should be independent of the order of the projectors up to a global phase, i.e, the parity of all elements in a projector should be the same.  The parity of the element $T_{\beta,\beta_1,\beta_2,\beta_3,\beta_4}[i,j]a^{\dag,\beta}_{(i,j)}
a_{(i,j)_l}^{\beta_1}a_{(i,j)_d}^{\beta_2}a_{(i,j)_r}^{\beta_3}a_{(i,j)_u}^{\beta_4}$ of the projector $\hat{P}[i,j]$ is obtained by $\tilde{p}(\beta)\tilde{p}(\beta_1)\tilde{p}(\beta_2)\tilde{p}(\beta_3)\tilde{p}(\beta_4)$, where
$\tilde{p}(x)$=-1, if the parity of $x$ is odd, and $\tilde{p}(x)$=+1 if the parity of $x$ is even.
Therefore, the parity of all elements can be obtained by the lower indices of tensor $T_{\beta,\beta_1,\beta_2,\beta_3, \beta_4}[i,j]$. Without lose of generality, we assume all nonzero projector elements have even parity in this paper.
As a consequence, the elements with odd parity vanish, i.e.,
$T_{\beta,\beta_1,\beta_2,\beta_3, \beta_4}[i,j]$=0, if $\beta$+$\beta_1$+$\beta_2$+$\beta_3$+$\beta_4$ is odd.
In this definition of fPEPS, we may interchange the positions of any two projectors and EPR pairs,
because they all have even parity.

One of the key issues in the fPEPS is the order of the fermion operators, including the
operators in the projectors and in EPRs.
%We need to specify the order of the virtual particles in the projectors and EPR pairs.
%According to the former definition of the fPEPS, the order of the virtual particles in the fEPR state and the order of the fermionic opertors in the projectors is still undefine.
We define the standard order of the fermion operators in each projector operators on the square lattice as followings, physical creation operator, left, down, right, and up virtual operators (i.e., anti-clockwise order), which is the same as the order of the lower indices in the tensor $T_{\beta,\beta_1,\beta_2,\beta_3,\beta_4}$
(see Eq.~\ref{eq:projector}). When changing the order of fermion operators, a sign which is determined by the parity of the indices will appear. For example, if we exchange the two
adjacent fermion operators $a_{(i,j),r}^{\beta_3}$ and $a_{(i,j),u}^{\beta_4}$, there will be an extra phase, i.e., $T_{\beta,\beta_1,\beta_2,\beta_4,\beta_3}=\tilde{p}(\beta_3,\beta_4)T_{\beta,\beta_1,\beta_2,\beta_3,\beta_4}$, where
\begin{eqnarray}\label{equ::phase}
\tilde{p}(\beta_3,\beta_4)&=&
\left\{
	\begin{array}{rll}
		-1,  &\ \ {\rm both} \ \ \tilde{p}(\beta_3),\ \ \tilde{p}(\beta_4)=-1, \\
		 1,  &\ \ {\rm otherwise}.
	\end{array}
\right.
\end{eqnarray}

\begin{figure} [tb]
		\begin{center}
		\includegraphics[width=0.3\textwidth]{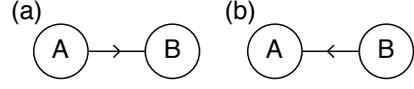}
		\caption{Schematic diagrams of Fermi arrows between tensor A and tensor B,
corresponding to (a) Eq.\ref{eq:lr} and (b) Eq.~\ref{eq:rl}.
}\label{pic:contract1}
		\end{center}
\end{figure}

Beside the fermion operators appeared in projector $\hat{P}$,
we also need to specify the operators' order in the EPR pairs, which is not given in the tensors explicitly.
%\red{\sout{In previous methods, \cite{Corboz2010,Gu2010,Kraus2010}
%the operator orders in EPS pairs are usually
%assumed beforehand and fixed during calculations. }}
%\red{\sout{These methods work well if the contraction order is fixed.
%%However, f
%For some algorithms, e.g., the GO method introduced in this paper,
%the fermion operators' order in an EPR may change during the calculations to improve the efficiency. In these cases, the orders of the EPR pairs have to be considered explicitly.}}
In this work, we introduce a {\it Fermi arrow} notation to specify the order of the EPR pairs.
For example, as shown in Fig.~\ref{pic:contract1}(a), the arrow points from site A to site B ,
and the corresponding EPR state is  $\hat{I}_{A\rightarrow B} |0\rangle=\sum^{D-1}_{k=0}a_B^{\dag k}a_A^{\dag k}|0\rangle$,
whereas in Fig.~\ref{pic:contract1}(b), the arrow points from B to A,
and the corresponding EPR state is $\hat{I}_{B\rightarrow A}|0\rangle=\sum^{D-1}_{k=0}a_A^{\dag k}a_B^{\dag k}|0\rangle $.
The two states can be transformed to each other as follows,
\begin{equation}
  \sum^{D-1}_{k=0}a_A^{\dag k}a_B^{\dag k}|0\rangle=\sum^{D-1}_{k=0}(-1)^{\tilde{p}(k)}a_B^{\dag k}a_A^{\dag k}|0\rangle \, .
\end{equation}

\begin{figure} [tb]
		\begin{center}
		\includegraphics[width=0.3\textwidth]{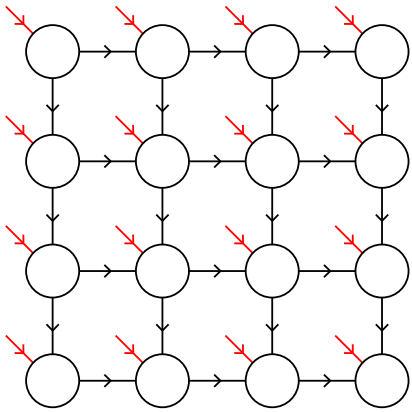}
		\caption{A schematic example of a fPEPS on the 4$\times$4 lattice. The  circles are the tensors
on the lattice whereas the black solid lines are the virtual EPR pairs and
the arrows on the bonds specify the order
of the fermion operators in the EPR pairs. The red solid lines are the physical indices associated with creation operators.
}
\label{pic:fPEPS}
		\end{center}
\end{figure}

We may also assign Fermi arrows to the physical indices: the Fermi arrows
point into the sites for the physical creation operators, and pointing out of the sites for the annihilation operators.
This definition is equivalent to insert EPR pairs between the physical operators when contracting the physical indices e.g.,
$\langle 0| \sum_{\beta_A,\beta_B}A^\dag_{\beta_A} a_A^{\beta_A} B_{\beta_B} a_B^{\dag,\beta_B}|0\rangle=\langle 0| \sum_{\beta_A}A^\dag_{\beta_A} a_A^{\beta_A} \sum_{\beta_B} B_{\beta_B} a_B^{\beta_B} \sum_k a_B^{\dag k}a_A^{\dag k}|0\rangle$. With this definition, we can treat the physical indices and the virtual indices in the same manner,
and do not need to distinguish the real fermions and virtual fermions during operations.
We can now uniquely define a fPEPS on a directed lattice,
as shown for example in Fig.~\ref{pic:fPEPS}, on a 4$\times$4 lattice.

%\subsection{Contraction}\label{sec:contraction}

By defining some calculation rules associated with Fermi arrows,
we are able to perform fPEPS calculations.
Contraction is one of the most important operations in PEPS algorithms.
When we contract the tensors on two sites in a fPEPS,
we need to consider the Fermi arrow direction.
We take the two situations in Fig.~\ref{pic:contract1} as an example,
which gives two different contraction formula,
\begin{equation}
\sum_{\beta,\beta_A,\beta_B}A_{\beta_A}a_A^{\beta_A} B_{\beta_B} a_B^{\beta_B}a_B^{\dag\beta}a_A^{\dag\beta}|0\rangle=\sum_{\beta} A_{\beta}B_{\beta}|0\rangle
\label{eq:lr}
\end{equation}
for Fig.~\ref{pic:contract1}(a) and,
\begin{equation}
\sum_{\beta,\beta_A,\beta_B}A_{\beta_A}a_A^{\beta_A} B_{\beta_B} a_B^{\beta_B}a_A^{\dag\beta}a_B^{\dag\beta}|0\rangle=\sum_\beta \tilde{p}(\beta)A_\beta B_\beta |0\rangle
\label{eq:rl}
\end{equation}
for Fig.~\ref{pic:contract1}(b).
The anti-commutation relation of fermions has been used.
The Fermi arrow helps to distinguish the two situations of contraction in
the graphical notions of Fig.~\ref{pic:contract1}.
Using the graphic representation may greatly simplify the notation.

More generally, giving two tensors $\mathbf{A}$ and $\mathbf{B}$, connected via multi virtual bonds (EPRs),
where $\{i_1,i_2,\cdots,i_k, i_{k+1}, \cdots, i_N\}$ are the joint bonds to be contracted. Assume that bonds $\{i_1,i_2,\cdots,i_k \}$
have Fermi arrows pointing from $\mathbf{B}$  to $\mathbf{A}$ , and the rest bonds $\{i_{k+1},\cdots i_{N}\}$
have Fermi arrows pointing from $\mathbf{A}$ to $\mathbf{B}$.
We first reshape $\mathbf{A}$ to $A_{I,i_N,i_{N-1},\cdots,i_1}$ and reshape $\mathbf{B}$ to $B_{i_1,i_2,\cdots,i_N,J}$,
where $\{I \}$, $\{J \}$ are the bonds that are not to be contracted in $\mathbf{A}$ and $\mathbf{B}$ respectively.
For the convenience of notation, we assume the signs due to the change of bond order in the tensors according to Eq.~\ref{equ::phase} have been absorbed into the tensors, then the resulting tensor is,
\begin{equation} R_{I,J}=\sum_{i_1,i_2,\cdots,i_N}\prod_{l=1,k}\tilde{p}(i_{l})A_{I,i_N,i_{N-1},\cdots,i_1}B_{i_1,i_2,\cdots,i_N,J}
\label{eq:contraction}
\end{equation}

Other often used operations associated with Fermi arrows are given in the Appendix.
We implement these operation rules for fPEPS in our recently developed TNSpack \cite{Dong2017libTNSP},
in which the anti-commutation properties of the fermion operators are automatically taken account of by these rules. Therefore, one can implement fPEPS algorithm very similar to that of standard PEPS without worrying too much
about the details of the anti-commutation between the fermion operators.

The Fermi arrows define the fermionic order for the fPEPS.
In some previous methods, \cite{Corboz2010}
the EPS pairs are not explicitly used. We note that in this work, the EPR pairs are only used in the derivation of the operation rules associated with Fermi arrows. Once we have these rules, one may ignore the underlying EPR pairs, and use only Fermi arrows for all operations.
In Ref.~\onlinecite{Barthel2009}, the authors proposed a general fermionization procedure using so called fermionic operator circuits (FOCs), in a bra and ket notation, instead of EPR pairs. Our Fermi arrows are similar to
the contraction arcs defined in Ref.~\onlinecite{Barthel2009} albeit the starting point and detailed implementations of the two methods are different.

\section{Stochastic gradient optimization of fPEPS}\label{GM}

In order to find the ground state of a given Hamiltonian using fPEPS, different methods have been introduced. The leading method is the imaginary time evolution (ITE) method.\cite{Xiang2008}
However, due to the high computation complexity to obtain the exact environment during the time evolution, some kinds of approximations are necessary. The simple update method\cite{Xiang2008} has been widely used, which however may have large
errors because the environment is over simplified. Several methods have been developed to treat the environment more rigorously, such as the full update method\cite{Lubasch2014,Michael2014}, and the gradient method\cite{Vanderstraeten2016,Liu2017}, which may significantly improve the results, but the scaling to $D$ of these methods is rather high.

We recently developed stochastic gradient optimization method for PEPS, combined with Monte Carlo sampling techniques. \cite{Liu2017,He2018} This method gives remarkable accuracy of the results which may be even better than the results of full update method at given $D$.\cite{Liu2017}
 The method has two advantages. First, the
environments of tensors are treated rigorously, and therefore, the results are more accurate than SU and even FU methods\cite{Liu2017}. Secondly, the Monte Carlo sampling technique may greatly reduce the scaling of the method to the virtual bond dimension $D$ from $D^{10}$ to $D^6$ for OBC,\cite{Sandvik2007, Schuch2008, Cirac2010, Banuls2017} which is even more crucial for fPEPS, where larger $D$ is often needed to converge the results. In this work, we extended this method to fPEPS.

The fPEPS wave functions of a many-particle system in Eq.~\ref{equ:fPEPS} can be rewritten as,
\begin{eqnarray}\label{MCTNS}
|\Psi\rangle&=&\sum_{\{n\}} \mathcal{C}\Big[\prod_{\bf i} T[{\bf i}]_{n,\beta_{1},\beta_{2},\beta_{3},\beta_{4}}\Big]|{n_1,n_2,\dots,n_N}\rangle \nonumber \\
     &\equiv & \sum_{\{\bf n\}} W({\bf n})|{\bf n}\rangle \, ,
\end{eqnarray}
where ${\bf i}$=$(i,j)$ is the site index of the lattice, and $\mathcal{C}$ means to contract all the entangled virtual fermions according to the rules defined in Sec.~\ref{sec:fpeps}.
$W({\bf n})$ is the coefficient of the physical state $|{\bf n}\rangle$= $|{n_1,n_2,\dots,n_N}\rangle$ in the particle number representation.
The energy of the system can be written as,
\begin{eqnarray}\label{MCEnergy}
	E&=&\frac{\langle \Psi|H|\Psi\rangle}{\langle \Psi|\Psi\rangle} \nonumber \\
	 &=& {1 \over \sum_{\bf n'} |W({\bf n}')|^2} \sum_{\bf n} |W({\bf n})|^2 E({\bf n})
\end{eqnarray}
where,
\begin{equation}
	E({\bf n})=\sum_{{\bf n}'}\frac{W({\bf n}')}{W({\bf n})}\langle {\bf n}'|H| \bf{n}\rangle \, .
\end{equation}
The total energy of the system for a given fPEPS can be evaluated via Monte Carlo sampling over the physical configurations space. \cite{Sandvik2007, Schuch2008, Cirac2010, Banuls2017, Liu2017}

To optimize the energy function, we need the derivatives of the energy with respect to the tensor elements,
\begin{eqnarray}
	\frac{\partial E}{\partial T[{\bf i}]_{n,\beta_{1},\beta_{2},\beta_{3},\beta_{4}}}
=2\langle\Delta[{\bf i}]_{n,\beta_{1},\beta_{2},\beta_{3},\beta_{4}}({\bf n})E({\bf n})\rangle\\ \nonumber
	-2\langle\Delta[{\bf i}]_{n,\beta_1,\beta_2,\beta_3,\beta_4}({\bf n})\rangle\langle E({\bf n})\rangle ,
\end{eqnarray}
where $\langle\cdots\rangle$ denotes the MC average. $\Delta[{\bf i}]_{n,\beta_1,\beta_2,\beta_3,\beta_4}$ is defined as
\begin{equation}
	\Delta[{\bf i}]_{n,\beta_1,\beta_2,\beta_3,\beta_4}(n)=\frac{1}{W({\bf n})}\frac{\partial W({\bf n})}{\partial T[{\bf i}]_{n,\beta_1,\beta_2,\beta_3,\beta_4}} ,
\end{equation}
and the derivative of $W({\bf n})$ is
\begin{equation}
	\frac{\partial W({\bf n})}{\partial T[{\bf i}]_{n,\beta_1,\beta_2,\beta_3,\beta_4}}=\mathcal{C}\Big[T[1]\cdots T[{\bf i-1}]T[{\bf i+1}]\cdots T[N]  \Big] \,.
\label{eq:delta}
\end{equation}
The derivatives can be also evaluated by the MC samplings.\cite{Liu2017}

Once we have the energy and its gradients, we can optimize the system energy using stochastic gradient method\cite{Sandvik2007,Liu2017}, which has been successfully applied to the standard PEPS method.

The overall algorithm for fPEPS is similar to that of PEPS. We need to contract the fPEPS
tensors at given particle configuration to obtain $W({\bf n})$ and the gradients. However, contracting
a fPEPS is much more complicate than contacting a standard PEPS, because of the anti-commutation relation of fermions. One must be very careful about the contraction order and underlying fermions' order in EPR pairs.
We show here that with the help of Fermi arrows and the operation rules associated with them, the contraction can be
done easily as in the standard PEPS algorithms.

\begin{figure} [tb]
		\begin{center}
		\includegraphics[width=0.3\textwidth]{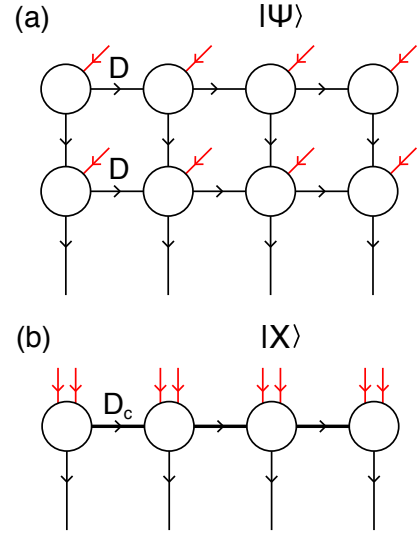}
		\caption{Approximate (a) a double row PEPS with bond dimension $D$ by (b) a MPO with bond dimension $D_c$.}
\label{pic:contract2}
		\end{center}
\end{figure}

\begin{figure} [tb]
		\begin{center}
		\includegraphics[width=0.45\textwidth]{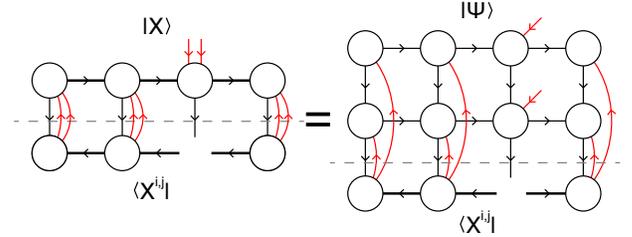}
		\caption{The equation $\langle X^{i,j}|X\rangle=\langle X^{i,j}|\Psi\rangle$, where $|\Psi\rangle$ is shown in Fig.~\ref{pic:contract2}(a) and $|X\rangle$ is shown in Fig.~\ref{pic:contract2}(b). $|X^{i,j}\rangle$ is obtained by taking the tensor $T^{i,j}$ out of $|X\rangle$.
}
\label{pic:equMPS}
		\end{center}
\end{figure}
\begin{figure} [tb]
		\begin{center}
		\includegraphics[width=0.3\textwidth]{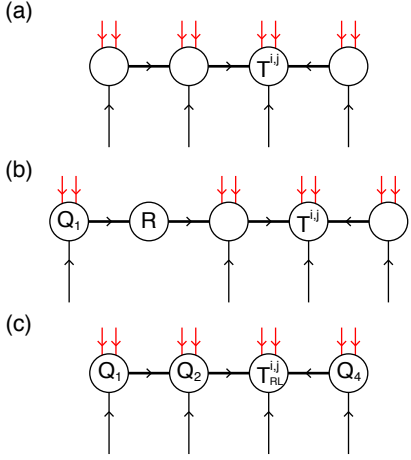}
		\caption{Simplification of $|X\rangle$: (a) To make use of the orthogonality conditions,
we first reverse the directions some Fermi arrows. (b) We perform QR decompositions starting form the first tensor.
(c) The tensor state $|X\rangle$ after simplification.}
\label{pic:LQQR}
		\end{center}\label{pic:LQQRdecomposition}
\end{figure}

To obtain $W({\bf n})$, we need to contact a single layer of fPEPS with fixed particle configuration $|{\bf n}\rangle$.
We adopt the boundary-MPO method, \cite{Lubasch2014,Michael2014}where we need to find a fermionic matrix product operator
(fMPO) denoted as $|X \rangle$ with bond dimension $D_c$ [see Fig.\ref{pic:contract2}(b)] to approximate the
two rows of fPEPS with bond dimension $D$, denoted as $|\Psi\rangle$ [see Fig.\ref{pic:contract2}(a)].
To find such $|X\rangle$, we minimize
\begin{equation}
	\delta=\frac{| |X \rangle - |\Psi\rangle | ^2}{\langle \Psi|\Psi\rangle}  \, ,
\label{eq:min}
\end{equation}
 which lead to the linear equation for each tensor $T^{i,j}$ on site $(i,j)$,
\begin{equation}
	\langle X^{i,j}|X\rangle=\langle X^{i,j}|\Psi\rangle \, ,
\label{eq:Tuv}
\end{equation}
where $|X^{i,j}\rangle$ is obtained by taking the tensor $T^{i,j}$ out of $|X\rangle$,
as graphically shown in Fig.~\ref{pic:equMPS}.

To solve the equation, we first contract the tensors on the left side of Fig.~\ref{pic:equMPS}.
We change the arrow directions from Fig.~\ref{pic:contract2}(b) to  Fig.~\ref{pic:LQQR}(a),
i.e., all arrows are pointing into site $(i,j)$.
The rule of changing the directions of Fermi arrows are given in the Appendix.
As will be seen in the following text, the change of Fermi arrow directions is to take the advantages of
the canonic form of fMPO.\cite{Garcia2006}

We next do QR decomposition to the tensor on the first site of  $|X\rangle$,
resulting in two tensors, $Q_1$ and $R$ as shown in Fig.~\ref{pic:LQQR}(b).
The rules for QR (and other decompositions) in the presence of Fermi arrows are
also given in the Appendix.
We then contract the $R$ tensor with the second tensor on the right site, and
perform QR decomposition on the second site again to obtain the $Q_2$ tensor. We repeat this process until reach
the tensor $T^{i,j}$. We contract the last $R$ tensor with the $T^{i,j}$, resulting in a new
tenor $T^{i,j}_R$.
Similarly, we perform the LQ decomposition on the right side of $|X\rangle$,
starting from the last site to the site ($i$, $j$), and
contract the last $L$ tensor with $T^{i,j}_R$ to get $T^{i,j}_{RL}$.
During the LQ (QR) decompositions, new Fermi arrows have been inserted between L (Q) tensors and Q (R) tensor.
After these processes, we obtain $|X\rangle$ in Fig.~\ref{pic:LQQR}(c).

\begin{figure} [tb]
		\begin{center}
		\includegraphics[width=0.4\textwidth]{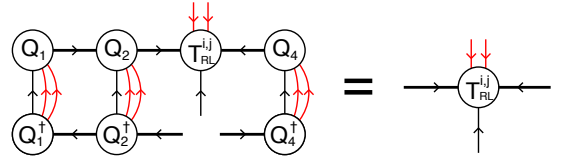}
		\caption{By applying the orthogonality conditions $Q_i Q_i^{\dag}$=${\bf I}$, the left side
of Fig.~\ref{pic:equMPS} reduces to a single tensor $T^{i,j}_{RL}$.
}\label{pic:LQQRequ}
		\end{center}
\end{figure}

\begin{figure} [tb]
		\begin{center}
		\includegraphics[width=0.4\textwidth]{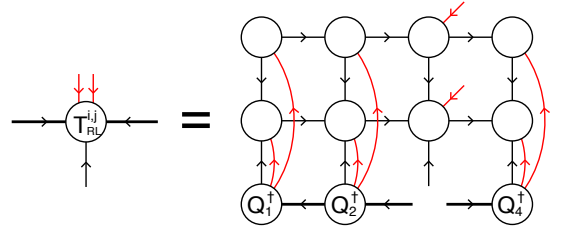}
		\caption{ The equation in Fig.~\ref{pic:equMPS} after the LQ and the QR decomposition on the $|X^{i,j}\rangle$. }\label{pic:equMPS2}
		\end{center}
\end{figure}

We perform the same operations for $|X^{i,j} \rangle$.
After these operations, the left side of Fig.~\ref{pic:equMPS} become that of Fig.~\ref{pic:LQQRequ}.
By using the orthogonality of $\hat{Q}$ and $\hat{Q}^\dag$, i.e., $\hat{Q}\hat{Q}^\dag$=${\bf I}$,
which is discussed in the Appendix for the fPEPS with Fermi arrows, we obtain the right side of Fig.~\ref{pic:LQQRequ}.
The original equation Fig.~\ref{pic:equMPS} become of Fig.~\ref{pic:equMPS2}, which can be solved iteratively
as in standard boundary MPO method,\cite{Lubasch2014,Michael2014} which usually converges in a few sweeps.

The contraction in Eq.~\ref{eq:delta} can be calculated in the same procedures.
Once we have  $W({\bf n})$,
and $\Delta_{n_i\beta_1\beta_2\beta_3\beta_4}$, the energies and their gradients can be easily calculated.

In our calculations, we first perform ITE with simple update method to obtain a approximate ground state,\cite{Liu2017}
which usually have energy errors around  $10^{-2}$.
We further optimize the fPEPS via gradient decent method to obtain the highly accurate ground state.

\section{benchmark results}

We benchmark our method for two typical fermion models on finite size square lattices, including
the interacting spin-less fermions model and the t-J model. We demonstrate that our method can give highly
accurate results compared with the exact results.

\subsection{The spin-less fermions model}

The interacting spin-less fermions model reads,
\begin{equation}
H=-t\sum_{\langle i,j\rangle}(c_i^\dag c_j + H.c.) + V\sum_{\langle i,j \rangle}n_i n_j \, ,
\end{equation}
where $c_i^\dag$ and $c_j$ are the creation and the annihilation operators, and $\langle i,j \rangle$ denotes
the nearest neighbor pairs. The chemical potential $\mu$ is set to zero here. We set hopping parameter $t$=1, and the interaction strength $V\geq 0$. In general this model is not exactly solvable, and has been numerically investigated in Ref.~\onlinecite{Woul2010} by the mean field theory and in Ref.~\onlinecite{Corboz2010} by the iPEPS method. Both methods give similar phase diagrams. For the parameters we used, the ground state is in a uniform metallic phase when $V$ is small and moves towards the phase boundary between the uniform phase and the phase separation with the increasing of $V$. Therefore the ground state of the model is expected to have entanglement beyond area law.~\cite{Wolf2006}

We firstly calculate this model on a 4$\times$4 square lattice so we can compare the fPEPS results with those obtained from the exact diagonlization method. In the calculations, we take $D=10$, and $D_c=30\sim 40$. The convergence of these parameters will be discussed in details in Sec.\ref{convergence}. The results are presented in Table~\ref{tab:interacting} for various $V$. As seen from Table I, the SU method may give the results with errors around 5$\times$10$^{-3}$ when $V$ is small, but the errors increase
for larger $V$. When $V$=2, the error of SU is about 10$^{-2}$. The GO may significantly improve the ground state energies.
By using the given $D$ and $D_c$, we are able to obtain an impressive highly accurate ground state with relative error near $\sim 10^{-5}$.

\begin{table} [t]
\caption{The ground state energies of the interacting spinless fermion models on a $4\times 4$ lattice.
The ground state energies obtained from the simple update (SU) method, and the gradient optimization (GO)
method are compared with those from exact diagonalization method. The relative errors between the
GO results and exact results are in the order of 10$^{-5}$.   }
\begin{tabular}{ c c  c c c c}
\hline
$V$ &SU & GO & Exact & relative error\\
\hline
\hline
0.1  &  -0.66590 &-0.67124 & -0.67125 & 1$\times$10$^{-5}$ \\
0.8  &  -0.59255 &-0.59309 &-0.59312 &  5$\times$10$^{-5}$  \\
2  & -0.48136 &-0.50643 & -0.50646 & 5$\times$10$^{-5}$ \\
\hline
\hline
\end{tabular}
\label{tab:interacting}
 \end{table}

We now consider a special case of $V$=0, where the model reduces to the free fermion model.
Although in this case, the model is exactly solvable,  it is a challenging model for the fPEPS method because
the free fermions have strong entanglement in real space $S\sim L^{d-1} {\rm log}L$ that violates the area law~\cite{Wolf2006}. Especially at $\mu$=0, the Fermi surface is very large, making the problem more
difficult. One may expect that to obtain the high accuracy results it requires very large $D$ and $D_c$. Furthermore, the required parameters $D$ and $D_c$ will generally increase rapidly with the increasing of the size of the system to keep the given accuracy.
In Table~\ref{tab:free}, we list the ground states energies of the free fermion model on the square lattice with different sizes obtained from the SU and GO methods, compared with the exact results $E_{ex}$.
We see that the relative errors of the SU are usually about $10^{-2}$ for $D$=8, but sometimes
the SU method may have numerical instability in some small systems.
The performance of GO is much better, and we always get stable results.
Even with a small $D$=6, the relative errors are about $10^{-3}\sim 10^{-4}$, and
reduce to $\sim$ $10^{-5}$ when $D$=8 is used. On the other hand, the violation of the area law is also showed in this table, that the accuracy gets lower in larger systems for a given $D$.

From the above tests, we find that the SU method sometimes may give rather accurate results ($\sim$ 10$^{-2}$ - 10$^{-3}$), but the situation may change from case to case. On the other hand the GO always gives reliable and highly accurate results ($\sim$ 10$^{-5}$).

\begin{table*} [t]
\caption{Compare the ground state energies of the free fermion model of SU ($D$=8), and GO ($D$=6, 8) with the exact results. For the 4$\times$4 lattice, the SU result is numerically instable for $D$=8. }
 \centering
 \begin{tabular}{ c c c c  c c c c c}
 \hline
 \mc{1}{c}{} & \mc{1}{c}{SU(D=8)} & \mc{3}{c}{GO(D=6)} & \mc{3}{c}{GO(D=8)} &  \\
  Size & Energy & Energy & $D_c$ & relative error & Energy & $D_c$ & relative error & exact\\
 \hline
 \hline
 4$\times$4   & -        & -0.68398 & 16 & 4$\times$10$^{-5}$ &  -0.68401 & 32  & 1$\times$10$^{-5}$ & -0.68402\\
 6$\times$6   & -0.67721 & -0.73269 & 24 & 5$\times$10$^{-4}$ &  -0.73305 & 52  & 5$\times$10$^{-5}$ & -0.73309\\
 8$\times$8   & -0.74763 & -0.75414 & 40 & 1$\times$10$^{-3}$ &  -0.75492 & 84  & 2$\times$10$^{-4}$ & -0.75510\\
 10$\times$10 & -0.75387 & -0.76619 & 55 & 1$\times$10$^{-3}$ &  -0.76705 & 110 & 5$\times$10$^{-4}$ & -0.76748\\
 12$\times$12 &     -    & -0.77094 & 80 & 5$\times$10$^{-3}$ &     -     & -   & -                & -0.77538 \\
\hline
 \hline
 \end{tabular}
\label{tab:free}
 \end{table*}

\subsection{t-J model}

In this section, we benchmark our method on the t-J model,
\begin{equation}
H=-t\sum_{\langle i,j\rangle,\sigma}(c_{i,\sigma}^\dag c_{j,\sigma} + H.c.) + J\sum_{\langle i,j\rangle}(\vec{S}_i\vec{S}_j-\frac{1}{4}n_i n_j)
\end{equation}
where $\sigma=\uparrow,\downarrow$ is the spin index and $\vec{S}_i$ is the spin $1/2$ operator on site $i$. $n_i=\sum_\sigma c_{i,\sigma}^\dag c_{i,\sigma}$
is the number of electrons on site $i$. In t-J model, the electron double occupancy is forbidden.
The t-J model is one of the key models to understand many important physical phenomena\cite{Zhang1988}, such as high $T_c$ superconductivity \cite{Lee2006}. Here, we calculate the model with $J/t=0.4$ and hole filling of $\bar{n}_h=$0.125.
The U(1) symmetry is adopted to enforce the particle number conservation.
But true physics of the system at this point, whether the ground state is a stripe state \cite{Hellberg1999,White1998,Corboz2011,Corboz2014} or an uniform phase\cite{Sherman2003,Vineet2018}, is still under debate.
Without doubt, the energy is one of the critical criterions to determine the ground state of the system.
We calculate the ground energies of different system sizes,
using $D=12$, $D_c=36 \sim 50$, and the results are shown in Table~\ref{tab:t-J} for both SU and GO methods. Again we see GO method greatly improves the energies obtained from SU method.
By extrapolating the energy to thermodynamic limit, we obtain that the ground state energy -0.6701,
which is lower than the value -0.6693 \cite{Corboz2014} from state of art DMRG calculations \cite{White1998} with $\chi \rightarrow \infty $, and -0.6619 obtained from variation quantum Monte Carlo plus $p$-step Lanczos methods.~\cite{Hu2012}
More results of t-J model~\cite{Dong_tj} will be published in a separate paper.

\begin{table} [t]
\caption{Compare the ground state energies of t-J model with hole filling $\bar{n}_h$=0.125
calculated by SU and GO methods. A virtual bond dimension $D=12$ is used. }
\begin{tabular}{  c c c }
\hline
size & SU & GO \\
\hline
\hline
4$\times$4   & -0.55108 &  -0.56420 \\
4$\times$8   & -0.57994 &  -0.59055 \\
6$\times$8   & -0.59431 &  -0.60349 \\
8$\times$8   & -0.60849 &  -0.61184 \\
8$\times$10  & -0.61068 &  -0.61738 \\
8$\times$12  & -0.61707 &  -0.62164 \\
12$\times$12 & -0.62307 &  -0.62973 \\
$L \rightarrow \infty$ &-0.66757 & -0.67008 \\
\hline
\hline
\end{tabular}
\label{tab:t-J}
 \end{table}

\section{convergence of fPEPS}\label{convergence}

Fermion systems may have large entanglement that beyond the area law \cite{Wolf2006} and
therefore it may need large $D$ to represent the many-particle state.
One may also expect that the $D$ and $D_c$ will increase with the size of the system.
The speed of the increasing of $D$ and $D_c$ along
with the size of the system indicates the efficiency of the simulation methods. It is important to understand how the fPEPS calculations converge respect to $D$ and $D_c$.
For finite systems, we can explicitly exam what $D$ and $D_c$ are needed to converge the results
as the size of the systems grow up.
In this section, we will discuss the convergence of the parameters $D$ and $D_c$ respectively.
We show that the behavior depends strongly on the models.

We first investigate
the convergency of the ground state energies to $D_c$ in a given system with fixed parameter $D$.
We calculate the error of energy defined as,
\begin{equation}\label{err}
\Delta E=E(D_c)-E(D_c^{\rm max}) \, ,
\end{equation}
where $E(D_c)$ is the energy with a giving truncation parameter $D_c$ and $E(D_c^{\rm max})$ is the converged energy
where the maximal $D_c$ is used.

\begin{figure} [tb]
		\begin{center}
        \includegraphics[width=0.4\textwidth]{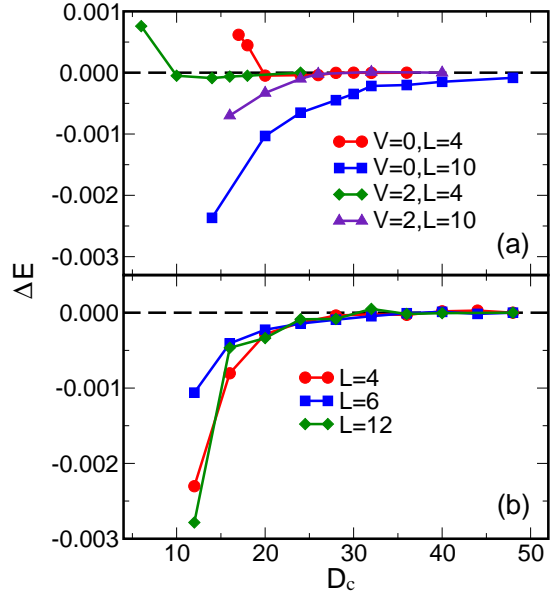}
		\caption{(a) The energy errors $\Delta E$ (See Eq.~\ref{err}) as functions of $D_c$ for the interacting electron model on the 4$\times$4 and 10$\times$10 lattices. A virtual bond dimension $D=8$ is used in the calculations. (b) The energy errors $\Delta E$  as functions of $D_c$ for the t-J model with $\bar{n}_h$=0.125 on the  4$\times$4, 6$\times$6 and 12$\times$12 lattices. A virtual bond dimension $D=$12 is used in the calculations. The dashed black line is a guide to the eye.
 }\label{pic:DeltaE_vs_Dc}
		\end{center}
\end{figure}

Figure~\ref{pic:DeltaE_vs_Dc}(a) depicts the results of
the spin-less fermion model with $V=0$ (free fermion) and $V=2$; and different system sizes, the $4\times4$ and $10\times 10$ lattices. We fix the bond dimension at $D=8$.
We use $D_c^{\rm max}=48$ for the $4\times 4$ system and $D_c^{\rm max}=64$ for the $10\times 10$ system.
We first note that $\Delta E$s approach 0 in a non-trivial way, which are not always from above (i.e., $\Delta E >$0).
This means that the ground state energy is not variational to $D_c$, and therefore one must be very careful
to extrapolate $D_c$ to infinite.
The convergency of energy is model dependent.
As shown in the figure, $D_c$ converge much faster for $V=2$ (correlated electrons) than for $V$=0 (free electrons).
In both cases, the convergency of energy strongly depend on the size of the systems.
In the cases of small sizes $L$=4, the energies converge rather fast with $D_c$.
However, for the $10\times 10$ system, $\Delta E$ converge much slower as functions of $D_c$. For $V=2$,
the energy is well converged at $D_c=26$ (about 3$D$), whereas the energy of free fermions
is not well converged even at $D_c=48$.

We investigate the convergence of the t-J model at hole doping $\bar{n}_h$=0.125, and the results are shown in Fig.~\ref {pic:DeltaE_vs_Dc}(b).
In the calculations, $D=12$ is used, and the result of $D_c^{\rm max}$=50 is used as a reference.
Interestingly, we find the ground state energies converge rather fast with $D_c$.
The errors reduce to 3$\times 10^{-4}$ for $D_c$=2$D$, and the errors reduce to 1$\times 10^{-5}$ for $D_c$=3$D$,
More importantly, unlike the interacting fermion model,
$D_c$ is only slightly dependent on the size of the system.

\begin{figure} [tb]
		\begin{center}
        \includegraphics[width=0.35\textwidth]{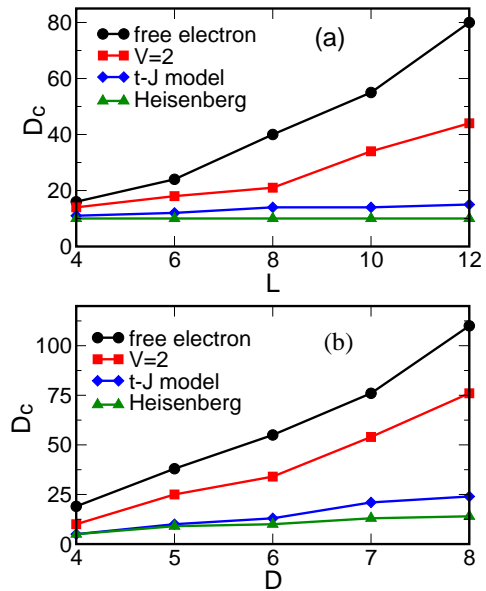}
		\caption{The bond truncation dimensions $D_c$ needed to ensure the contraction error $\delta<$10$^{-6}$  (see Eq.~\ref{eq:min}) as functions of (a) the lattice size $L$ and (b) the virtual bond dimension $D$ for various models,
including the spin-less interacting fermion models, and the t-J model. The ``t-J model'' in the figure is calculated with parameters $J=0.4$, $\bar{n}_h$=0.125, whereas the ``Heisenberg'' model is calculated using t-J model
in the limit of $\bar{n}_h$=0.
}\label{pic:Dc}
		\end{center}
\end{figure}

The energy errors in the calculations are induced by the contraction errors due
to bond dimension truncation.
We further test the relationship between the convergent truncation dimension $D_c$ and the size of the system, i.e. we exam the minimal $D_c$ needed to ensure the relative contracting error $\delta < 10^{-6}$ (see Eq.~\ref{eq:min}) along with the increasing of the system size. The bond dimension used here is fixed to a relatively small one $D=6$.
We compare the truncation errors for the spin-less interacting electron model at $V$=0 and $V$=2.
For the t-J model, we compare two situations, the hole doping $\bar{n}_h$=0.125, and the $\bar{n}_h$=0,
and the later one reduces to the Heisenberg model.
The results are shown in Fig.~\ref{pic:Dc}(a).
We find that the required $D_c$ is almost independent of the size of the system for the Heisenberg model, and for the t-J model with hole filling $\bar{n}_h$=0.125.
However, the required $D_c$ increases rapidly with the size of the system for the interacting electron model, especially for the  free electrons. At $L$=12, $D_c$=80 is required to ensure the desired contraction accuracy
for the free electron model  and $D_c$=40 for the $V$=2 model.

We also exam the relationship between $D_c$ and the bond dimension $D$. In this test, we fixed the size of the system
to $L$=10. The results are shown in Fig.~\ref{pic:Dc}(b). We see that the required $D_c$ increase roughly linearly with $D$
for these models. For the Heisenberg model (and even J$_1$-J$_2$ model) \cite{Liu2018} and the t-J model with hole filling $\bar{n}_h=$0.125, $D_c$$\sim$ 2$D$ - 3$D$ is enough to ensure the accuracy of contraction, whereas for the interacting electron model, $D_c$$\sim$ 9$D$ - 15$D$ are required to ensure the desired contraction accuracy, which becomes the major difficult to simulate these models.
We note that in the standard contraction method, the bond truncation dimension $D_{c_2}$ for a double layer tensor network should scale as $D^2$,~\cite{Lubasch2014,Michael2014} making the simulation of fermions with large $D$ even more difficult.

\begin{figure} [tb]
		\begin{center}
        \includegraphics[width=0.35\textwidth]{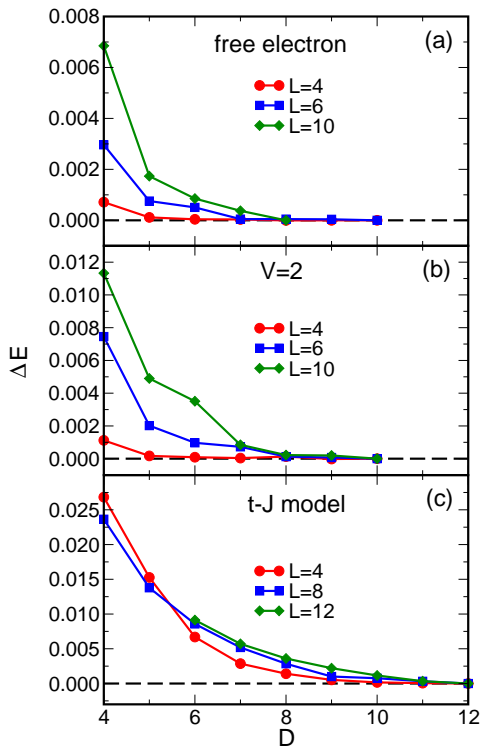}
		\caption{The convergence of ground state energy as functions of D for (a) the free fermion model, (b) the interacting spin-less fermions model with $V$=2,  and (c) the t-J model with $\bar{n}_h$=0.125, on the lattices of different sizes. The dashed black line is a guide to the eye.}\label{pic:convegerD}
		\end{center}
\end{figure}

With the convergent $D_c$ for each $D$, we can analyze the convergence of the energy against $D$ for a given system size. The energy of a model in the thermodynamic limit can be further extracted by finite size scaling method. The convergence of the energy against the parameter $D$ are shown in Fig.~\ref{pic:convegerD}, where $\Delta E$ is defined as the energy differences compared to those of maxima $D$, which are $D^{\rm max}$=8 for the free fermion model and $D^{\rm max}$=10 for the interacting fermion model with $V$=2.  A $D^{\rm max}$=12 is used for the t-J model with $\bar{n}_h$=0.125.
Surprisingly, for the free fermion model ($V$=0), we have $\Delta E \approx 4\times 10^{-4}$ for $D$=7 on the 10$\times$10 lattice, as shown in Fig.~\ref{pic:convegerD}(a), where one may expect a much larger error. For the interacting fermion with $V=2$, which is shown in Fig.~\ref{pic:convegerD}(b), the energy is also converged to $\Delta E \approx 1\times 10^{-4}$ at $D$=8. On the other hand, for the t-J model, the energies converge rather slowly with $D$. For the $12 \times 12$ system, the energy errors reduces to about $3\times 10^{-4}$ at $D=11$.
The non-trivial dependence of $D$ and $D_c$ for different models may pose some interesting questions to understand the structure of fPEPS. We leave these problems for future studies.

\section{Summary}

In this work, we extend the stochastic gradient optimization method combined
with Monte Carlo sampling techniques
to optimize the fPEPS wave functions for fermion systems.
The Monte Carlo sampling techniques may greatly reduce the scaling of the
calculation, and therefore allow using larger bond dimensions ($D$)
and bond truncation dimensions ($D_c$) in the calculations,
which is important for the faithful simulations of fermion systems.

We benchmark the method on the interacting spinless fermion models,
and the t-J models. The numerical calculations show that the gradient optimization
may greatly enhance the accuracy of the results over the simple update method.
We further investigate the converge of fPEPS calculation with respect
to $D$ and $D_c$ for the models. The free fermion model is most challenging
to simulate with fPEPS, because the $D_c$ increase very rapidly with $D$ and the
size of the system. For t-J models, we find that large $D$s are needed to converge the results.
Our method therefore offer a powerful tool to simulate fermion systems
because it has much lower scaling in both computational time and memory
than direct contraction methods.

\section{Acknowledge}
This work was funded by the National Key Research and Development Program of China (Grant No. 2016YFB0201202), the Chinese National Science Foundation (Grants No. 11774327, No. 11874343, No. 11474267), and the Strategic Priority Research Program (B) of the Chinese Academy of Sciences (Grant No. XDB01030200). China Postdoctoral Science Foundation funded project (Grant No. 2018M632529). The numerical calculations have been done on the USTC HPC facilities.

\appendix

\section{Rules for Fermi arrows}

In this Appendix, we give the rules of operations associated with the
Fermi arrows in fPEPS. These rules are straightforward to prove.

\subsubsection{Reversing Fermi arrows and the Hermitian conjugate}
\label{sec:reverse}

\begin{figure} [tb]
		\begin{center}
		\includegraphics[width=0.3\textwidth]{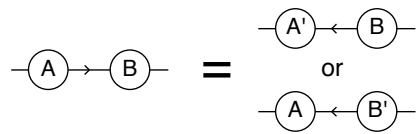}
		\caption{The rule for reversing the Fermi arrow. }\label{pic:reverse}
		\end{center}
\end{figure}

Sometimes, we need to reverse the direction of a Fermi arrow. The rule of reversing Fermi arrows is
giving as follows. Suppose,
\begin{eqnarray}
\hat{A} &=&\sum_{\beta_1,\beta_2} A_{\beta_1,\beta_2}a_1^{\beta_1} a_2^{\beta_2} \, ,\nonumber\\
\hat{B} &=&\sum_{\beta_3,\beta_4} B_{\beta_3,\beta_4}a_3^{\beta_3} a_4^{\beta_4} \, , \nonumber
\end{eqnarray}
are two projectors in a fPEPS that are connected by a Fermi arrow pointing from
$\hat{A}$ to $\hat{B}$, as shown on the left side of Fig.~\ref{pic:reverse}.
We may reverse the Fermi arrow, pointing from $\hat{B}$ to $\hat{A}$, and resulting in
two possible (but equivalent) forms that are given on the right side of Fig.~\ref{pic:reverse}.
It is easy to prove that,
\begin{eqnarray}
\hat{A'}&=&\sum_{\beta_1,\beta_2}\tilde{p}(\beta_2)A_{\beta_1,\beta_2}a_1^{\beta_1} a_2^{\beta_2} , \nonumber \\
\hat{B'}&=&\sum_{\beta_3,\beta_4}\tilde{p}(\beta_3)B_{\beta_3,\beta_4}a_3^{\beta_3} a_4^{\beta_4} .
\end{eqnarray}

\begin{figure} [tb]
		\begin{center}
		\includegraphics[width=0.3\textwidth]{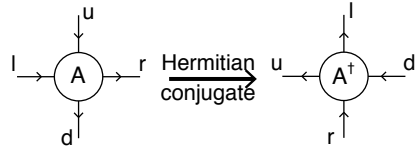}
		\caption{The rule of taking a Hermitian conjugate a tensor.}\label{pic:HC}
		\end{center}
\end{figure}

When we calculate the expectation value of a physical quantity, $\langle \Psi_{\rm fPEPS} | \hat{O} |\Psi_{\rm fPEPS} \rangle$, we need take the Hermitian conjugate of
ket state $|\Psi_{\rm fPEPS} \rangle$ to get the bra state $\langle \Psi_{\rm fPEPS} |$.
When taking the Hermitian conjugate of the projectors in a fPEPS,
we need to (i) reserve the orders of the indices of the tensor associated with the projectors,
e.g., change tensor $A_{l,d,r,u}$ to $A^{\dag}_{u,r,d,l}$,  as shown in Fig.~\ref{pic:HC};
and (ii) reverse all the Fermi arrows associated with the projectors.
Note that here the reversion of the Ferim arrows is required by the Hermitian conjugate, and no change is needed
for the tensors during the process.

\subsubsection{Matrix decompositions and contractions}\label{sec:SVD}

\begin{figure} [tb]
		\begin{center}
		\includegraphics[width=0.4\textwidth]{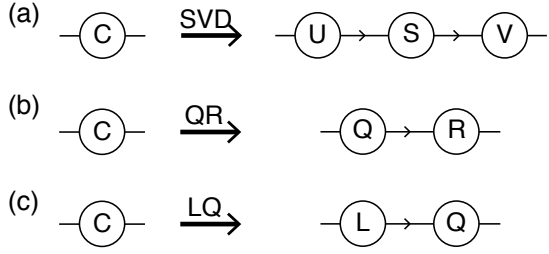}
		\caption{ The rules for (a) SVD decomposition, $\hat{C}=\hat{U}\hat{S}\hat{V} \hat{I}_{\hat{U}\rightarrow \hat{S}} \hat{I}_{\hat{S}\rightarrow \hat{V}}$; (b) the QR decomposition, $\hat{C}=\hat{Q}\hat{R} \hat{I}_{\hat{Q}\rightarrow \hat{R}} $; and (c) the LQ decomposition, $\hat{C}=\hat{L}\hat{Q}\hat{I}_{\hat{L}\rightarrow \hat{Q}}$. The EPR pairs with Fermi arrows has been inserted into decomposed matrices.
The bonds on the left and right side can have arrows in either direction, which keep unchanged after the decompositions.}\label{pic:SVDandConjugate}
		\end{center}
\end{figure}

The operations such as tensor decompositions also have close relation to the Fermi arrows.
For example, in the standard PEPS, when we do SVD to a matrix $C$, we have $C$=$USV$.
However, in fPEPS, two Fermi arrows should be inserted to the inner bonds after the decomposition, i.e.,
the Fermi arrow pointing from $\hat{U}$ to $\hat{S}$, and the one pointing from $\hat{S}$ to $\hat{V}$
as follows, and schematically shown in Fig.~\ref{pic:SVDandConjugate}(a),
\begin{eqnarray}\label{equ:SVD}
%\hat{C}=\hat{U}\hat{S}\hat{V}|I_{f:\hat{U}\rightarrow \hat{S}}\rangle|I_{f:\hat{S}\rightarrow \hat{V}}\rangle,
\hat{C}=\hat{U}\hat{S}\hat{V} \hat{I}_{\hat{U}\rightarrow \hat{S}} \hat{I}_{\hat{S}\rightarrow \hat{V}},
\end{eqnarray}
where
\begin{eqnarray}
\hat{C}&=&\sum_{\alpha,\beta}C_{\alpha,\beta}a_{L}^{\alpha} a_R^{\beta} \nonumber \\
\hat{U}&=&\sum_{\alpha,\delta_U}U_{\alpha,\delta_U}a_{L}^{\alpha}a_{U}^{\delta_U} \nonumber\\
\hat{S}&=&\sum_{\delta_1,\delta_2}S_{\delta_1,\delta_2}a_{S_1}^{\delta_1}a_{S_2}^{\delta_2} \nonumber\\
\hat{V}&=&\sum_{\delta_V,\beta}V_{\delta_V,\beta}a_{V}^{\delta_V}a_{R}^{\beta}
\end{eqnarray}
Other matrix decompositions such as LQ/QR decompositions follow the similar rules, i.e., one need to
insert Fermi arrows (i.e., directed EPR pairs) between the decomposed matrices, as shown in Fig.~\ref{pic:SVDandConjugate}(b),(c).

\begin{figure} [htb]
		\begin{center}
		\includegraphics[width=0.3\textwidth]{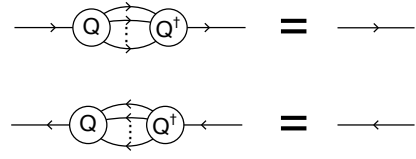}
		\caption{The rule of contacting $\hat{Q}_i \hat{Q}_i^{\dag}$. To use the orthogonality $\hat{Q}_i \hat{Q}_i^{\dag}$=${\bf I}$, the Fermi arrows must have ``consistent directions'' as shown above.
}\label{pic:Orthogonality}
		\end{center}
\end{figure}

In standard PEPS, we often use so called canonical form of MPS in the MPO algorithm\cite{Lubasch2014,Michael2014} to contract the PEPS, taking the advantage of the orthogonality of the $Q_i$ tensors obtained from LQ/QR decompositions (or the $U$ and $V$ matrices
from SVD decompostions),~\cite{Schollwoeck2011}
i.e.,  $Q_i Q_i^{\dag}$=${\bf I}$, where ${\bf I}$ is a unit matrix.
However, this relation cann't be directly used in the fPEPS, where we need to take the Fermi arrows into consideration
during the contractions. It is easily prove that only when the Fermi arrows have ``consistent directions'', i.e., all Fermi arrows point from $\hat{Q}$ to $\hat{Q}^{\dag}$, or from $\hat{Q}^{\dag}$ to $\hat{Q}$, we can use the orthogonality condition for $\hat{Q}$ matrix. The results after contraction are Fermi arrows pointing to the right or to the left,
as schematically shown in Fig.~\ref{pic:Orthogonality}. If the Fermi arrows are not ``consistent'',
we need to rearrange the directions of the Fermi arrows first to make them ``consistent'',
before we can use the orthogonality condition. This is done in Sec.\ref{GM}, when we contract two rows of
fPEPS via a MPO scheme.

%
%\bibliographystyle{apsrev}
%\bibliography{MyBib}

\begin{thebibliography}{47}
\expandafter\ifx\csname natexlab\endcsname\relax\def\natexlab#1{#1}\fi
\expandafter\ifx\csname bibnamefont\endcsname\relax
  \def\bibnamefont#1{#1}\fi
\expandafter\ifx\csname bibfnamefont\endcsname\relax
  \def\bibfnamefont#1{#1}\fi
\expandafter\ifx\csname citenamefont\endcsname\relax
  \def\citenamefont#1{#1}\fi
\expandafter\ifx\csname url\endcsname\relax
  \def\url#1{\texttt{#1}}\fi
\expandafter\ifx\csname urlprefix\endcsname\relax\def\urlprefix{URL }\fi
\providecommand{\bibinfo}[2]{#2}
\providecommand{\eprint}[2][]{\url{#2}}

\bibitem[{\citenamefont{Lee et~al.}(2006)\citenamefont{Lee, Nagaosa, and
  Wen}}]{Lee2006}
\bibinfo{author}{\bibfnamefont{P.~A.} \bibnamefont{Lee}},
  \bibinfo{author}{\bibfnamefont{N.}~\bibnamefont{Nagaosa}}, \bibnamefont{and}
  \bibinfo{author}{\bibfnamefont{X.-G.} \bibnamefont{Wen}},
  \bibinfo{journal}{Rev. Mod. Phys.} \textbf{\bibinfo{volume}{78}},
  \bibinfo{pages}{17} (\bibinfo{year}{2006}).

\bibitem[{\citenamefont{Stormer et~al.}(1999)\citenamefont{Stormer, Tsui, and
  Gossard}}]{Stormer1999}
\bibinfo{author}{\bibfnamefont{H.~L.} \bibnamefont{Stormer}},
  \bibinfo{author}{\bibfnamefont{D.~C.} \bibnamefont{Tsui}}, \bibnamefont{and}
  \bibinfo{author}{\bibfnamefont{A.~C.} \bibnamefont{Gossard}},
  \bibinfo{journal}{Rev. Mod. Phys.} \textbf{\bibinfo{volume}{71}},
  \bibinfo{pages}{S298} (\bibinfo{year}{1999}).

\bibitem[{\citenamefont{Edwards and Hewson}(1968)}]{EDWARDS1968}
\bibinfo{author}{\bibfnamefont{D.~M.} \bibnamefont{Edwards}} \bibnamefont{and}
  \bibinfo{author}{\bibfnamefont{A.~C.} \bibnamefont{Hewson}},
  \bibinfo{journal}{Rev. Mod. Phys.} \textbf{\bibinfo{volume}{40}},
  \bibinfo{pages}{810} (\bibinfo{year}{1968}).

\bibitem[{\citenamefont{Imada et~al.}(1998)\citenamefont{Imada, Fujimori, and
  Tokura}}]{Imada1998}
\bibinfo{author}{\bibfnamefont{M.}~\bibnamefont{Imada}},
  \bibinfo{author}{\bibfnamefont{A.}~\bibnamefont{Fujimori}}, \bibnamefont{and}
  \bibinfo{author}{\bibfnamefont{Y.}~\bibnamefont{Tokura}},
  \bibinfo{journal}{Rev. Mod. Phys.} \textbf{\bibinfo{volume}{70}},
  \bibinfo{pages}{1039} (\bibinfo{year}{1998}).

\bibitem[{\citenamefont{Foulkes et~al.}(2001)\citenamefont{Foulkes, Mitas,
  Needs, and Rajagopal}}]{Foulkes2001}
\bibinfo{author}{\bibfnamefont{W.~M.~C.} \bibnamefont{Foulkes}},
  \bibinfo{author}{\bibfnamefont{L.}~\bibnamefont{Mitas}},
  \bibinfo{author}{\bibfnamefont{R.~J.} \bibnamefont{Needs}}, \bibnamefont{and}
  \bibinfo{author}{\bibfnamefont{G.}~\bibnamefont{Rajagopal}},
  \bibinfo{journal}{Rev. Mod. Phys.} \textbf{\bibinfo{volume}{73}},
  \bibinfo{pages}{33} (\bibinfo{year}{2001}).

\bibitem[{\citenamefont{Li and Yao}(2018)}]{Li2018}
\bibinfo{author}{\bibfnamefont{Z.-X.} \bibnamefont{Li}} \bibnamefont{and}
  \bibinfo{author}{\bibfnamefont{H.}~\bibnamefont{Yao}},
  \bibinfo{journal}{arXiv:1805.08219v2}  (\bibinfo{year}{2018}).

\bibitem[{\citenamefont{Loh et~al.}(1990)\citenamefont{Loh, Gubernatis,
  Scalettar, White, Scalapino, and Sugar}}]{Loh1990}
\bibinfo{author}{\bibfnamefont{E.~Y.} \bibnamefont{Loh}},
  \bibinfo{author}{\bibfnamefont{J.~E.} \bibnamefont{Gubernatis}},
  \bibinfo{author}{\bibfnamefont{R.~T.} \bibnamefont{Scalettar}},
  \bibinfo{author}{\bibfnamefont{S.~R.} \bibnamefont{White}},
  \bibinfo{author}{\bibfnamefont{D.~J.} \bibnamefont{Scalapino}},
  \bibnamefont{and} \bibinfo{author}{\bibfnamefont{R.~L.} \bibnamefont{Sugar}},
  \bibinfo{journal}{Phys. Rev. B} \textbf{\bibinfo{volume}{41}},
  \bibinfo{pages}{9301} (\bibinfo{year}{1990}).

\bibitem[{\citenamefont{Troyer and Wiese}(2005)}]{Troyer05}
\bibinfo{author}{\bibfnamefont{M.}~\bibnamefont{Troyer}} \bibnamefont{and}
  \bibinfo{author}{\bibfnamefont{U.-J.} \bibnamefont{Wiese}},
  \bibinfo{journal}{Phys. Rev. Lett.} \textbf{\bibinfo{volume}{94}},
  \bibinfo{pages}{170201} (\bibinfo{year}{2005}).

\bibitem[{\citenamefont{Schollwöck}(2011)}]{Schollwoeck2011}
\bibinfo{author}{\bibfnamefont{U.}~\bibnamefont{Schollwöck}},
  \bibinfo{journal}{Annals of Physics} \textbf{\bibinfo{volume}{326}},
  \bibinfo{pages}{96 } (\bibinfo{year}{2011}), \bibinfo{note}{january 2011
  Special Issue}.

\bibitem[{\citenamefont{Perez-Garcia et~al.}(2007)\citenamefont{Perez-Garcia,
  Verstraete, Wolf, and Cirac}}]{Garcia2006}
\bibinfo{author}{\bibfnamefont{D.}~\bibnamefont{Perez-Garcia}},
  \bibinfo{author}{\bibfnamefont{F.}~\bibnamefont{Verstraete}},
  \bibinfo{author}{\bibfnamefont{M.~M.} \bibnamefont{Wolf}}, \bibnamefont{and}
  \bibinfo{author}{\bibfnamefont{J.~I.} \bibnamefont{Cirac}},
  \bibinfo{journal}{Quantum Info. Comput.} \textbf{\bibinfo{volume}{7}},
  \bibinfo{pages}{401} (\bibinfo{year}{2007}).

\bibitem[{\citenamefont{Verstraete et~al.}(2008)\citenamefont{Verstraete, Murg,
  and Cirac}}]{Verstraete2008}
\bibinfo{author}{\bibfnamefont{F.}~\bibnamefont{Verstraete}},
  \bibinfo{author}{\bibfnamefont{V.}~\bibnamefont{Murg}}, \bibnamefont{and}
  \bibinfo{author}{\bibfnamefont{J.}~\bibnamefont{Cirac}},
  \bibinfo{journal}{Advances in Physics} \textbf{\bibinfo{volume}{57}},
  \bibinfo{pages}{143} (\bibinfo{year}{2008}).

\bibitem[{\citenamefont{Jiang et~al.}(2008)\citenamefont{Jiang, Weng, and
  Xiang}}]{Xiang2008}
\bibinfo{author}{\bibfnamefont{H.~C.} \bibnamefont{Jiang}},
  \bibinfo{author}{\bibfnamefont{Z.~Y.} \bibnamefont{Weng}}, \bibnamefont{and}
  \bibinfo{author}{\bibfnamefont{T.}~\bibnamefont{Xiang}},
  \bibinfo{journal}{Phys. Rev. Lett.} \textbf{\bibinfo{volume}{101}},
  \bibinfo{pages}{090603} (\bibinfo{year}{2008}).

\bibitem[{\citenamefont{Vidal}(2008)}]{Vidal2008}
\bibinfo{author}{\bibfnamefont{G.}~\bibnamefont{Vidal}},
  \bibinfo{journal}{Phys. Rev. Lett.} \textbf{\bibinfo{volume}{101}},
  \bibinfo{pages}{110501} (\bibinfo{year}{2008}).

\bibitem[{\citenamefont{{Verstraete} and {Cirac}}(2004)}]{Verstraete2004}
\bibinfo{author}{\bibfnamefont{F.}~\bibnamefont{{Verstraete}}}
  \bibnamefont{and} \bibinfo{author}{\bibfnamefont{J.~I.}
  \bibnamefont{{Cirac}}}, \bibinfo{journal}{arXiv:cond-mat/0407066}
  (\bibinfo{year}{2004}).

\bibitem[{\citenamefont{Sfondrini et~al.}(2010)\citenamefont{Sfondrini,
  Cerrillo, Schuch, and Cirac}}]{Cirac2010}
\bibinfo{author}{\bibfnamefont{A.}~\bibnamefont{Sfondrini}},
  \bibinfo{author}{\bibfnamefont{J.}~\bibnamefont{Cerrillo}},
  \bibinfo{author}{\bibfnamefont{N.}~\bibnamefont{Schuch}}, \bibnamefont{and}
  \bibinfo{author}{\bibfnamefont{J.~I.} \bibnamefont{Cirac}},
  \bibinfo{journal}{Phys. Rev. B} \textbf{\bibinfo{volume}{81}},
  \bibinfo{pages}{214426} (\bibinfo{year}{2010}).

\bibitem[{\citenamefont{Verstraete et~al.}(2006)\citenamefont{Verstraete, Wolf,
  Perez-Garcia, and Cirac}}]{Verstraete06}
\bibinfo{author}{\bibfnamefont{F.}~\bibnamefont{Verstraete}},
  \bibinfo{author}{\bibfnamefont{M.~M.} \bibnamefont{Wolf}},
  \bibinfo{author}{\bibfnamefont{D.}~\bibnamefont{Perez-Garcia}},
  \bibnamefont{and} \bibinfo{author}{\bibfnamefont{J.~I.} \bibnamefont{Cirac}},
  \bibinfo{journal}{Phys. Rev. Lett.} \textbf{\bibinfo{volume}{96}},
  \bibinfo{pages}{220601} (\bibinfo{year}{2006}).

\bibitem[{\citenamefont{Wang et~al.}(2016)\citenamefont{Wang, Gu, Verstraete,
  and Wen}}]{wang16}
\bibinfo{author}{\bibfnamefont{L.}~\bibnamefont{Wang}},
  \bibinfo{author}{\bibfnamefont{Z.-C.} \bibnamefont{Gu}},
  \bibinfo{author}{\bibfnamefont{F.}~\bibnamefont{Verstraete}},
  \bibnamefont{and} \bibinfo{author}{\bibfnamefont{X.-G.} \bibnamefont{Wen}},
  \bibinfo{journal}{Phys. Rev. B} \textbf{\bibinfo{volume}{94}},
  \bibinfo{pages}{075143} (\bibinfo{year}{2016}).

\bibitem[{\citenamefont{Vidal}(2007)}]{Vidal2007}
\bibinfo{author}{\bibfnamefont{G.}~\bibnamefont{Vidal}},
  \bibinfo{journal}{Phys. Rev. Lett.} \textbf{\bibinfo{volume}{98}},
  \bibinfo{pages}{070201} (\bibinfo{year}{2007}).

\bibitem[{\citenamefont{M.-H. et~al.}(2012)\citenamefont{M.-H., Cirac, and
  Bañuls}}]{Alexander2012}
\bibinfo{author}{\bibfnamefont{A.}~\bibnamefont{M.-H.}},
  \bibinfo{author}{\bibfnamefont{J.~I.} \bibnamefont{Cirac}}, \bibnamefont{and}
  \bibinfo{author}{\bibfnamefont{M.~C.} \bibnamefont{Bañuls}},
  \bibinfo{journal}{New Journal of Physics} \textbf{\bibinfo{volume}{14}},
  \bibinfo{pages}{075003} (\bibinfo{year}{2012}).

\bibitem[{\citenamefont{Barthel et~al.}(2009)\citenamefont{Barthel, Pineda, and
  Eisert}}]{Barthel2009}
\bibinfo{author}{\bibfnamefont{T.}~\bibnamefont{Barthel}},
  \bibinfo{author}{\bibfnamefont{C.}~\bibnamefont{Pineda}}, \bibnamefont{and}
  \bibinfo{author}{\bibfnamefont{J.}~\bibnamefont{Eisert}},
  \bibinfo{journal}{Phys. Rev. A} \textbf{\bibinfo{volume}{80}},
  \bibinfo{pages}{042333} (\bibinfo{year}{2009}).

\bibitem[{\citenamefont{Corboz et~al.}(2010)\citenamefont{Corboz, Or\'us,
  Bauer, and Vidal}}]{Corboz2010}
\bibinfo{author}{\bibfnamefont{P.}~\bibnamefont{Corboz}},
  \bibinfo{author}{\bibfnamefont{R.}~\bibnamefont{Or\'us}},
  \bibinfo{author}{\bibfnamefont{B.}~\bibnamefont{Bauer}}, \bibnamefont{and}
  \bibinfo{author}{\bibfnamefont{G.}~\bibnamefont{Vidal}},
  \bibinfo{journal}{Phys. Rev. B} \textbf{\bibinfo{volume}{81}},
  \bibinfo{pages}{165104} (\bibinfo{year}{2010}).

\bibitem[{\citenamefont{Corboz et~al.}(2011)\citenamefont{Corboz, White, Vidal,
  and Troyer}}]{Corboz2011}
\bibinfo{author}{\bibfnamefont{P.}~\bibnamefont{Corboz}},
  \bibinfo{author}{\bibfnamefont{S.~R.} \bibnamefont{White}},
  \bibinfo{author}{\bibfnamefont{G.}~\bibnamefont{Vidal}}, \bibnamefont{and}
  \bibinfo{author}{\bibfnamefont{M.}~\bibnamefont{Troyer}},
  \bibinfo{journal}{Phys. Rev. B} \textbf{\bibinfo{volume}{84}},
  \bibinfo{pages}{041108} (\bibinfo{year}{2011}).

\bibitem[{\citenamefont{{Gu} et~al.}(2010)\citenamefont{{Gu}, {Verstraete}, and
  {Wen}}}]{Gu2010}
\bibinfo{author}{\bibfnamefont{Z.-C.} \bibnamefont{{Gu}}},
  \bibinfo{author}{\bibfnamefont{F.}~\bibnamefont{{Verstraete}}},
  \bibnamefont{and} \bibinfo{author}{\bibfnamefont{X.-G.} \bibnamefont{{Wen}}},
  \bibinfo{journal}{arXiv:1004.2563}  (\bibinfo{year}{2010}).

\bibitem[{\citenamefont{Gu et~al.}(2013)\citenamefont{Gu, Jiang, Sheng, Yao,
  Balents, and Wen}}]{Gu2013}
\bibinfo{author}{\bibfnamefont{Z.-C.} \bibnamefont{Gu}},
  \bibinfo{author}{\bibfnamefont{H.-C.} \bibnamefont{Jiang}},
  \bibinfo{author}{\bibfnamefont{D.~N.} \bibnamefont{Sheng}},
  \bibinfo{author}{\bibfnamefont{H.}~\bibnamefont{Yao}},
  \bibinfo{author}{\bibfnamefont{L.}~\bibnamefont{Balents}}, \bibnamefont{and}
  \bibinfo{author}{\bibfnamefont{X.-G.} \bibnamefont{Wen}},
  \bibinfo{journal}{Phys. Rev. B} \textbf{\bibinfo{volume}{88}},
  \bibinfo{pages}{155112} (\bibinfo{year}{2013}).

\bibitem[{\citenamefont{Kraus et~al.}(2010)\citenamefont{Kraus, Schuch,
  Verstraete, and Cirac}}]{Kraus2010}
\bibinfo{author}{\bibfnamefont{C.~V.} \bibnamefont{Kraus}},
  \bibinfo{author}{\bibfnamefont{N.}~\bibnamefont{Schuch}},
  \bibinfo{author}{\bibfnamefont{F.}~\bibnamefont{Verstraete}},
  \bibnamefont{and} \bibinfo{author}{\bibfnamefont{J.~I.} \bibnamefont{Cirac}},
  \bibinfo{journal}{Phys. Rev. A} \textbf{\bibinfo{volume}{81}},
  \bibinfo{pages}{052338} (\bibinfo{year}{2010}).

\bibitem[{\citenamefont{Hastings}(2007)}]{Hastings2007}
\bibinfo{author}{\bibfnamefont{M.~B.} \bibnamefont{Hastings}},
  \bibinfo{journal}{Journal of Statistical Mechanics: Theory and Experiment}
  \textbf{\bibinfo{volume}{2007}}, \bibinfo{pages}{P08024}
  (\bibinfo{year}{2007}).

\bibitem[{\citenamefont{Wolf}(2006)}]{Wolf2006}
\bibinfo{author}{\bibfnamefont{M.~M.} \bibnamefont{Wolf}},
  \bibinfo{journal}{Phys. Rev. Lett.} \textbf{\bibinfo{volume}{96}},
  \bibinfo{pages}{010404} (\bibinfo{year}{2006}).

\bibitem[{\citenamefont{Lubasch
  et~al.}(2014{\natexlab{a}})\citenamefont{Lubasch, Cirac, and
  Ba\~nuls}}]{Lubasch2014}
\bibinfo{author}{\bibfnamefont{M.}~\bibnamefont{Lubasch}},
  \bibinfo{author}{\bibfnamefont{J.~I.} \bibnamefont{Cirac}}, \bibnamefont{and}
  \bibinfo{author}{\bibfnamefont{M.-C.} \bibnamefont{Ba\~nuls}},
  \bibinfo{journal}{Phys. Rev. B} \textbf{\bibinfo{volume}{90}},
  \bibinfo{pages}{064425} (\bibinfo{year}{2014}{\natexlab{a}}).

\bibitem[{\citenamefont{Lubasch
  et~al.}(2014{\natexlab{b}})\citenamefont{Lubasch, Cirac, and
  Bañuls}}]{Michael2014}
\bibinfo{author}{\bibfnamefont{M.}~\bibnamefont{Lubasch}},
  \bibinfo{author}{\bibfnamefont{J.~I.} \bibnamefont{Cirac}}, \bibnamefont{and}
  \bibinfo{author}{\bibfnamefont{M.-C.} \bibnamefont{Bañuls}},
  \bibinfo{journal}{New Journal of Physics} \textbf{\bibinfo{volume}{16}},
  \bibinfo{pages}{033014} (\bibinfo{year}{2014}{\natexlab{b}}).

\bibitem[{\citenamefont{Corboz et~al.}(2014)\citenamefont{Corboz, Rice, and
  Troyer}}]{Corboz2014}
\bibinfo{author}{\bibfnamefont{P.}~\bibnamefont{Corboz}},
  \bibinfo{author}{\bibfnamefont{T.~M.} \bibnamefont{Rice}}, \bibnamefont{and}
  \bibinfo{author}{\bibfnamefont{M.}~\bibnamefont{Troyer}},
  \bibinfo{journal}{Phys. Rev. Lett.} \textbf{\bibinfo{volume}{113}},
  \bibinfo{pages}{046402} (\bibinfo{year}{2014}).

\bibitem[{\citenamefont{Sandvik and Vidal}(2007)}]{Sandvik2007}
\bibinfo{author}{\bibfnamefont{A.~W.} \bibnamefont{Sandvik}} \bibnamefont{and}
  \bibinfo{author}{\bibfnamefont{G.}~\bibnamefont{Vidal}},
  \bibinfo{journal}{Phys. Rev. Lett.} \textbf{\bibinfo{volume}{99}},
  \bibinfo{pages}{220602} (\bibinfo{year}{2007}).

\bibitem[{\citenamefont{Schuch et~al.}(2008)\citenamefont{Schuch, Wolf,
  Verstraete, and Cirac}}]{Schuch2008}
\bibinfo{author}{\bibfnamefont{N.}~\bibnamefont{Schuch}},
  \bibinfo{author}{\bibfnamefont{M.~M.} \bibnamefont{Wolf}},
  \bibinfo{author}{\bibfnamefont{F.}~\bibnamefont{Verstraete}},
  \bibnamefont{and} \bibinfo{author}{\bibfnamefont{J.~I.} \bibnamefont{Cirac}},
  \bibinfo{journal}{Phys. Rev. Lett.} \textbf{\bibinfo{volume}{100}},
  \bibinfo{pages}{040501} (\bibinfo{year}{2008}).

\bibitem[{\citenamefont{{Ba{\~n}uls} et~al.}(2017)\citenamefont{{Ba{\~n}uls},
  {Cichy}, {Ignacio Cirac}, {Jansen}, {K{\"u}hn}, and {Saito}}}]{Banuls2017}
\bibinfo{author}{\bibfnamefont{M.~C.} \bibnamefont{{Ba{\~n}uls}}},
  \bibinfo{author}{\bibfnamefont{K.}~\bibnamefont{{Cichy}}},
  \bibinfo{author}{\bibfnamefont{J.}~\bibnamefont{{Ignacio Cirac}}},
  \bibinfo{author}{\bibfnamefont{K.}~\bibnamefont{{Jansen}}},
  \bibinfo{author}{\bibfnamefont{S.}~\bibnamefont{{K{\"u}hn}}},
  \bibnamefont{and} \bibinfo{author}{\bibfnamefont{H.}~\bibnamefont{{Saito}}},
  in \emph{\bibinfo{booktitle}{European Physical Journal Web of Conferences}}
  (\bibinfo{year}{2017}), vol. \bibinfo{volume}{137}, p.
  \bibinfo{pages}{04001}.

\bibitem[{\citenamefont{Dong et~al.}(2017)\citenamefont{Dong, Liu, Zhou, Guo,
  Zhou, Han, and He}}]{Dong2017}
\bibinfo{author}{\bibfnamefont{S.-J.} \bibnamefont{Dong}},
  \bibinfo{author}{\bibfnamefont{W.}~\bibnamefont{Liu}},
  \bibinfo{author}{\bibfnamefont{X.-F.} \bibnamefont{Zhou}},
  \bibinfo{author}{\bibfnamefont{G.-C.} \bibnamefont{Guo}},
  \bibinfo{author}{\bibfnamefont{Z.-W.} \bibnamefont{Zhou}},
  \bibinfo{author}{\bibfnamefont{Y.-J.} \bibnamefont{Han}}, \bibnamefont{and}
  \bibinfo{author}{\bibfnamefont{L.}~\bibnamefont{He}}, \bibinfo{journal}{Phys.
  Rev. B} \textbf{\bibinfo{volume}{96}}, \bibinfo{pages}{045119}
  (\bibinfo{year}{2017}).

\bibitem[{\citenamefont{Liu et~al.}(2017)\citenamefont{Liu, Dong, Han, Guo, and
  He}}]{Liu2017}
\bibinfo{author}{\bibfnamefont{W.-Y.} \bibnamefont{Liu}},
  \bibinfo{author}{\bibfnamefont{S.-J.} \bibnamefont{Dong}},
  \bibinfo{author}{\bibfnamefont{Y.-J.} \bibnamefont{Han}},
  \bibinfo{author}{\bibfnamefont{G.-C.} \bibnamefont{Guo}}, \bibnamefont{and}
  \bibinfo{author}{\bibfnamefont{L.}~\bibnamefont{He}}, \bibinfo{journal}{Phys.
  Rev. B} \textbf{\bibinfo{volume}{95}}, \bibinfo{pages}{195154}
  (\bibinfo{year}{2017}).

\bibitem[{\citenamefont{He et~al.}(2018)\citenamefont{He, An, Yang, Wang, Chen,
  Wang, Liang, Dong, Sun, Han et~al.}}]{He2018}
\bibinfo{author}{\bibfnamefont{L.}~\bibnamefont{He}},
  \bibinfo{author}{\bibfnamefont{H.}~\bibnamefont{An}},
  \bibinfo{author}{\bibfnamefont{C.}~\bibnamefont{Yang}},
  \bibinfo{author}{\bibfnamefont{F.}~\bibnamefont{Wang}},
  \bibinfo{author}{\bibfnamefont{J.}~\bibnamefont{Chen}},
  \bibinfo{author}{\bibfnamefont{C.}~\bibnamefont{Wang}},
  \bibinfo{author}{\bibfnamefont{W.}~\bibnamefont{Liang}},
  \bibinfo{author}{\bibfnamefont{S.}~\bibnamefont{Dong}},
  \bibinfo{author}{\bibfnamefont{Q.}~\bibnamefont{Sun}},
  \bibinfo{author}{\bibfnamefont{W.}~\bibnamefont{Han}}, \bibnamefont{et~al.},
  \bibinfo{journal}{IEEE Transactions on Parallel and Distributed Systems}
  \textbf{\bibinfo{volume}{29}}, \bibinfo{pages}{2838} (\bibinfo{year}{2018}).

\bibitem[{\citenamefont{Dong et~al.}(2018)\citenamefont{Dong, Liu, Wang, Han,
  Guo, and He}}]{Dong2017libTNSP}
\bibinfo{author}{\bibfnamefont{S.-J.} \bibnamefont{Dong}},
  \bibinfo{author}{\bibfnamefont{W.-Y.} \bibnamefont{Liu}},
  \bibinfo{author}{\bibfnamefont{C.}~\bibnamefont{Wang}},
  \bibinfo{author}{\bibfnamefont{Y.}~\bibnamefont{Han}},
  \bibinfo{author}{\bibfnamefont{G.-C.} \bibnamefont{Guo}}, \bibnamefont{and}
  \bibinfo{author}{\bibfnamefont{L.}~\bibnamefont{He}},
  \bibinfo{journal}{Computer Physics Communications}
  \textbf{\bibinfo{volume}{228}}, \bibinfo{pages}{163 } (\bibinfo{year}{2018}).

\bibitem[{\citenamefont{Vanderstraeten
  et~al.}(2016)\citenamefont{Vanderstraeten, Haegeman, Corboz, and
  Verstraete}}]{Vanderstraeten2016}
\bibinfo{author}{\bibfnamefont{L.}~\bibnamefont{Vanderstraeten}},
  \bibinfo{author}{\bibfnamefont{J.}~\bibnamefont{Haegeman}},
  \bibinfo{author}{\bibfnamefont{P.}~\bibnamefont{Corboz}}, \bibnamefont{and}
  \bibinfo{author}{\bibfnamefont{F.}~\bibnamefont{Verstraete}},
  \bibinfo{journal}{Phys. Rev. B} \textbf{\bibinfo{volume}{94}},
  \bibinfo{pages}{155123} (\bibinfo{year}{2016}).

\bibitem[{\citenamefont{de~Woul and Langmann}(2010)}]{Woul2010}
\bibinfo{author}{\bibfnamefont{J.}~\bibnamefont{de~Woul}} \bibnamefont{and}
  \bibinfo{author}{\bibfnamefont{E.}~\bibnamefont{Langmann}},
  \bibinfo{journal}{Journal of Statistical Physics}
  \textbf{\bibinfo{volume}{139}}, \bibinfo{pages}{1033} (\bibinfo{year}{2010}).

\bibitem[{\citenamefont{Zhang and Rice}(1988)}]{Zhang1988}
\bibinfo{author}{\bibfnamefont{F.~C.} \bibnamefont{Zhang}} \bibnamefont{and}
  \bibinfo{author}{\bibfnamefont{T.~M.} \bibnamefont{Rice}},
  \bibinfo{journal}{Phys. Rev. B} \textbf{\bibinfo{volume}{37}},
  \bibinfo{pages}{3759} (\bibinfo{year}{1988}).

\bibitem[{\citenamefont{Hellberg and Manousakis}(1999)}]{Hellberg1999}
\bibinfo{author}{\bibfnamefont{C.~S.} \bibnamefont{Hellberg}} \bibnamefont{and}
  \bibinfo{author}{\bibfnamefont{E.}~\bibnamefont{Manousakis}},
  \bibinfo{journal}{Phys. Rev. Lett.} \textbf{\bibinfo{volume}{83}},
  \bibinfo{pages}{132} (\bibinfo{year}{1999}).

\bibitem[{\citenamefont{White and Scalapino}(1998)}]{White1998}
\bibinfo{author}{\bibfnamefont{S.~R.} \bibnamefont{White}} \bibnamefont{and}
  \bibinfo{author}{\bibfnamefont{D.~J.} \bibnamefont{Scalapino}},
  \bibinfo{journal}{Phys. Rev. Lett.} \textbf{\bibinfo{volume}{80}},
  \bibinfo{pages}{1272} (\bibinfo{year}{1998}).

\bibitem[{\citenamefont{{Sherman} and {Schreiber}}(2003)}]{Sherman2003}
\bibinfo{author}{\bibfnamefont{A.}~\bibnamefont{{Sherman}}} \bibnamefont{and}
  \bibinfo{author}{\bibfnamefont{M.}~\bibnamefont{{Schreiber}}},
  \bibinfo{journal}{European Physical Journal B} \textbf{\bibinfo{volume}{32}},
  \bibinfo{pages}{203} (\bibinfo{year}{2003}), \eprint{cond-mat/0302356}.

\bibitem[{\citenamefont{{Vineet Mallik} et~al.}(2018)\citenamefont{{Vineet
  Mallik}, {Gupta}, {Shenoy}, and {Krishnamurthy}}}]{Vineet2018}
\bibinfo{author}{\bibfnamefont{A.}~\bibnamefont{{Vineet Mallik}}},
  \bibinfo{author}{\bibfnamefont{G.~K.} \bibnamefont{{Gupta}}},
  \bibinfo{author}{\bibfnamefont{V.~B.} \bibnamefont{{Shenoy}}},
  \bibnamefont{and} \bibinfo{author}{\bibfnamefont{H.~R.}
  \bibnamefont{{Krishnamurthy}}}, \bibinfo{journal}{arXiv:1805.02429}
  (\bibinfo{year}{2018}).

\bibitem[{\citenamefont{Hu et~al.}(2012)\citenamefont{Hu, Becca, and
  Sorella}}]{Hu2012}
\bibinfo{author}{\bibfnamefont{W.-J.} \bibnamefont{Hu}},
  \bibinfo{author}{\bibfnamefont{F.}~\bibnamefont{Becca}}, \bibnamefont{and}
  \bibinfo{author}{\bibfnamefont{S.}~\bibnamefont{Sorella}},
  \bibinfo{journal}{Phys. Rev. B} \textbf{\bibinfo{volume}{85}},
  \bibinfo{pages}{081110} (\bibinfo{year}{2012}).

\bibitem[{Don()}]{Dong_tj}
\bibinfo{note}{S.-J. Dong, C. Wang, Y. Han, C. Yang, and L. He, {\it
  unpublished}.}

\bibitem[{\citenamefont{Liu et~al.}(2018)\citenamefont{Liu, Dong, Wang, Han,
  An, Guo, and He}}]{Liu2018}
\bibinfo{author}{\bibfnamefont{W.-Y.} \bibnamefont{Liu}},
  \bibinfo{author}{\bibfnamefont{S.}~\bibnamefont{Dong}},
  \bibinfo{author}{\bibfnamefont{C.}~\bibnamefont{Wang}},
  \bibinfo{author}{\bibfnamefont{Y.}~\bibnamefont{Han}},
  \bibinfo{author}{\bibfnamefont{H.}~\bibnamefont{An}},
  \bibinfo{author}{\bibfnamefont{G.-C.} \bibnamefont{Guo}}, \bibnamefont{and}
  \bibinfo{author}{\bibfnamefont{L.}~\bibnamefont{He}}, \bibinfo{journal}{Phys.
  Rev. B} \textbf{\bibinfo{volume}{98}}, \bibinfo{pages}{241109}
  (\bibinfo{year}{2018}).

\end{thebibliography}

\end{document}